\documentclass{emulateapj}
\shorttitle{Detection of Low Mass-Ratio Stellar Binary Systems}
\shortauthors{Gullikson et al.}
\usepackage{natbib}
\usepackage{graphicx}
\usepackage{longtable}
\usepackage{url}
\usepackage{multirow}
\usepackage{float}

\usepackage{supertabular}
\begin{document}
%%%%%%%%%%%%%%%%%%%%%%%%%%%%%%%%%%%%%%%% 
% NEW COMMANDS
% 
\newcommand{\msun}{\ensuremath{\,\rm M_{\odot}}\xspace}  % solar mass units of a number
\newcommand{\lsun}{\ensuremath{\,\rm L_{\odot}}\xspace}  % solar luminosity units of a number
\newcommand{\rsun}{\ensuremath{\,\rm R_{\odot}}\xspace}  % solar radii units of a number
\newcommand{\zsun}{\ensuremath{\,\rm Z_{\odot}}\xspace}        
\newcommand{\hh}{\ensuremath{\mathrm{H}_2}\xspace}     % molecular hydrogen

%%%%%%%%%%%%%%%%%%%%%%%%%%%%%%%%%%%%%%%% 
\providecommand{\e}[1]{\ensuremath{\times 10^{#1}}}

\submitted{Accepted to the Astrophysical Journal on October 19, 2012}
\title{Detection of Low Mass-Ratio Stellar Binary Systems}
\author{Kevin Gullikson\altaffilmark{1},
             Sarah Dodson-Robinson\altaffilmark{1}}

\altaffiltext{1}{Astronomy Department, University of Texas, 1 University
Station C1400, Austin, TX 78712, USA. \texttt{kgulliks@astro.as.utexas.edu}}

%\maketitle

\begin{abstract}
O- and B-type stars are often found in binary systems, but the low binary mass-ratio regime is relatively unexplored due to observational difficulties. Binary systems with low mass-ratios may have formed through fragmentation of the circumstellar disk rather than molecular cloud core fragmentation. We describe a new technique sensitive to G- and K-type companions to early B stars, a mass-ratio of roughly 0.1, using high-resolution, high signal-to-noise spectra. We apply this technique to a sample of archived VLT/CRIRES observations of nearby B-stars in the CO bandhead near 2300 nm. While there are no unambiguous binary detections in our sample, we identify HIP 92855 and HIP 26713 as binary candidates warranting follow-up observations. We use our non-detections to determine upper limits to the frequency of FGK stars orbiting early B-type primaries. 
\end{abstract}


\maketitle

\newpage
\section{Introduction}
O- and B-type stars are often found in binary or
multiple systems: \cite{Mason2009} estimate that at least $57\%$ of O-stars are
in spectroscopic multiple systems, and at least $75\%$ are in any type of binary or multiple system.
Yet the multiplicity fraction of high-mass stars may be underestimated due to the difficulty of detecting low-mass secondary stars \citep{Sana2011}.  While the mass-ratio distribution
is reasonably well known for high-mass binaries with mass-ratio $q \equiv M_s/M_p > 0.2$, there are almost no
constraints for low mass-ratio binaries. However, binaries of low mass-ratio are important probes of star formation since they may have
formed in a different way than approximately equal-mass binaries. Here we define low mass-ratios to be those with $q < 0.2$, where $M_s$ is the mass of the secondary (lower mass binary component), and $M_p$ is the mass of the primary (higher mass binary component). We also use the term ``low-mass'' to describe star with $M \lesssim 1 M_{\odot}$, and ''high-mass'' to describe stars with $M \gtrsim 5 M_{\odot}$

\subsection{Binary Formation in High-Mass Stars}
\label{sec:formation}
With such a high fraction of high-mass stars found in binary or multiple
systems, any theory of high-mass star formation should be able to
explain the high binary formation rate. There are several mechanisms by
which binary stars may form: fission \citep{Lyttleton1953,
  Lebovitz1974, Lebovitz1984}, in which a molecular core begins
spinning fast enough as it collapses that it splits into two stars;
core fragmentation \citep[see e.g.][]{Boss1979, Boss1986, Bate1995}, in which a collapsing core develops two or more
overdensities which then begin collapsing separately; and disk
fragmentation \citep[see e.g.][]{Kratter2006, Stamatellos2011}, in which the circumstellar disk surrounding the primary
star becomes gravitationally unstable and creates a secondary
star. While the fission scenario was once thought to be important, it
has since fallen out of favor because the viscous dissipation timescale, which would drive a spinning body towards fission, is much longer than the core collapse timescale \citep{Tohline2002} and because hydrodynamic simulations fail to cause
the rotating core to actually split rather than just deform
\citep{Tohline2001}. Both core and disk fragmentation are still
thought to be viable binary formation mechanisms. It is likely that
both mechanisms play a role in shaping the binary mass-ratio and
separation distributions. In the formation of higher-order multiples, it is even possible that both mechanisms operate in the same system.

The primary method of forming binary systems is thought to be core fragmentation. As a molecular cloud
begins isothermally collapsing, its density increases, causing the
Jeans mass to decrease. Thus, an initially Jeans-mass collapsing core
can fragment into smaller objects. Core fragmentation will initially yield binaries with separations
$10 AU < a < 1000 AU$, which may move closer by interacting
with the surrounding gas, a circumbinary disk, or through dynamical
interactions with other nearby stars \citep{BBB2002}. For $\sim 1
M_{\odot}$ primary stars, the observed mass-ratio distribution is well fit by assuming
both components of the binary are chosen randomly from the stellar
initial mass function, and later evolve through
accretion and dynamical interactions \citep{Kroupa2003}. Assuming
independent component masses chosen from the IMF, a binary with a $10 M_{\odot}$ primary would
most often have a $0.1 M_{\odot}$ secondary, giving an initial mass
ratio near $q=0.01$. The unmodified mass-ratio distribution of
high-mass binaries would therefore strongly favor low mass
ratios. However, accretion of high specific angular-momentum gas
from either the collapsing molecular core or
the circumbinary disk will preferentially be captured by the
lower-mass companion, driving the binary mass-ratio toward unity and
decreasing the orbital separation
\citep{Bate2000, BonnellBate2005}. Additionally, dynamical
interactions tend to replace low-mass binary companions with higher-mass ones, or to kick the lower-mass component out to a wide orbit and create a hierarchical triple system. Over time, these processes tend to create high-mass binary systems 
with nearly equal masses and small separations \citep{BBB2002}. Dynamical interactions are most important in dense environments where the probability of stellar encounters is high. Therefore, they are probably more important in dense OB star clusters than in the much looser OB associations, where many B stars are found.

Another potentially important way to form binary systems is disk instability
\citep[see e.g.][]{Kratter2006, Stamatellos2011}. In
this scenario, the fragment forms in an unstable circumstellar disk
with an initial separation of $\sim 100 AU$ and initial mass-ratios
near $q \sim 0.03$, similar to the core fragmentation case
\citep{Kratter2006}. The final mass-ratio is expected to rise, but not
as significantly as in the core fragmentation scenario in which the
fragment can form sooner. Typical mass-ratios near $q \sim 0.1$ are expected, though the picture is far from complete. 
A semi-analytical treatment of embedded protostellar disks by
\cite{Kratter2008} finds that massive stars with $M > 2M_{\odot}$ maintain $0.01 - 0.1
M_{\odot}$ in orbiting fragments after about 2 Myr. \cite{Krumholz2007}
simulate a $100 M_{\odot}$ collapsing core for a much shorter time (20 kyr),
but also find that the disk fragments and that the final fragment mass
ratio is $q \approx 0.1$. However, \cite{Krumholz2009} simulate the same mass core
but start it with a slow solid body rotation instead of a turbulent velocity field and run the model for about twice as long (57 kyr), and find that it leads to a very massive binary ($M_1 + M_2 = 70 \rm M_{\odot}$) with 
a mass-ratio $q = 0.7$. Work by \cite{Clarke2009} indicates that as
mass is transported inwards onto the star, the outer disk can become unstable at late times. This
instability can lead to a delayed disk fragmentation,
with a fragment mass-ratio in the range $0.1 < q <
0.5$. The simulations by \cite{Clarke2009} were done for a $\sim
1M_{\odot}$ primary star, and the delayed fragmentation occured after
about $10^5$ years. Delayed fragmentation may not be possible in disks surrounding
high-mass stars, as the time at which fragmentation occurs is
comparable to the disk dispersal timescale \citep{Klahr2006}. However,
if it does occur, the similar fragmentation and dispersal timescales suggest that the fragment
would not undergo significant accretion or migration and would leave a
wide binary ($a \approx 100 AU$) with mass-ratio in the range $0.1 < q <
0.5$.

Unfortunately, there are no true binary population synthethis
simulations for high-mass binary systems formed by either mechanism
discussed above. The lack of population synthethis models is driven
by computational issues; a collapsing
cloud that reproduces the stellar mass function \emph{and} generates
enough high-mass stars to meaningfully analyze the binary statistics
would have to be very massive and therefore difficult to
simulate. Disk fragmentation simulations often either stop the
simulation once a fragment forms rather than follow its mass accretion
history \citep[e.g.][]{Boss2011, Krumholz2007}, or lack high enough resolution to follow the secondary very
near the star \citep[e.g.][]{BonnellBate2005}. Fortunately, such models may be in the near future. Realistic simulations of massive collapsing molecular clouds have begun appearing that can meaningfully discuss the multiplicity of low-mass binary systems
\citep{Bate2012, Krumholz2012}. These simulations reproduce the observed increase in multiplicity fraction with primary star mass, but do not yet generate enough high-mass binary system to compare the parameter distributions to observations. Despite the current lack of models, we can draw the general conclusion that disk
fragmentation tends to produce lower-mass companions than core
fragmentation. For this reason, probing the low mass-ratio regime can
provide information on the relative importance of both scenarios in
forming high-mass binary systems, and may help constrain models once
computational power increases.

In addition to binary star formation, disk instability is often invoked as a way
to form planets of a few Jupiter masses orbiting $\sim 1 M_{\odot}$ stars.
While the massive star formation process as a
whole may not simply be a scaled up version of low-mass star
formation \citep{Zinnecker2007}, the process of disk fragmentation may
be. One expects that disks around high-mass stars, with correspondingly higher accretion rates
and more mass, fragment more often than disks around low-mass stars \citep{Boss2011, Boss2006,
  Sally2009, Kratter2006}. Thus, if disk fragmentation plays an important role
in high-mass star formation, it may also play a role in low-mass star
formation by creating $\sim 10M_{Jup}$ planets and substellar companions.

\subsection{Observing low mass-ratio binaries}
\label{sec:otherobs}
Detection of OB-star binaries with mass-ratio $q \approx 0.1$ or lower is very
difficult, since the ratio of the secondary flux $F_s$ to the primary
flux $F_p$ is $F_s/F_p \sim 10^{-3}$ or lower in the V-band. Imaging surveys
can detect such contrast ratios for wide orbits, but lose sensitivity
as the separation decreases below about $1^{\prime\prime}$
\citep[e.g.][]{Maiz2010}. Spectroscopic binary surveys do well for
short-period systems where a full orbit can be mapped in a reasonable
amount of time, but lose sensitivity for periods greater than about one year
\citep[e.g.][]{Sana2009, Evans2010}. However, low-mass companions ($q \lesssim
0.2$), which induce a small reflex motion on the primary, are very difficult to find with
traditional spectroscopic surveys.

One method to find low-mass companions to late B-type primaries is to
search for high x-ray emission. Stars later than about B3 are not
expected to have strong enough winds to emit X-rays
\citep{Gagne1997}, and stars of earlier type than about A7 do not have
a radiative-convective boundary that can drive a magnetic dynamo and
create an X-ray generating corona \citep{Schmitt1997}. Stars in
between spectral types B4 and A7 with strong X-ray emission are
thought to have young low-mass companions, because the luminosity and
X-ray spectral energy distribution is similar to observed T-Tauri
stars \citep{Huelamo2000}. \cite{Evans2011} use this fact to search fo
low-mass companions to late B-stars in the open cluster Trumpler
16. They find a significant number of 
companions, and set the multiplicity fraction at $39\%$. This value is
a lower limit, but the authors believe that the true value is not much
above $39\%$.  Unfortunately, X-ray imaging is not effective for
primary star spectral types earlier than B3, which are also strong
x-ray emitters \citep{Gagne1997} and will drown out any companions.

\begin{figure*}[t]
  \centering
  \includegraphics[width=\columnwidth]{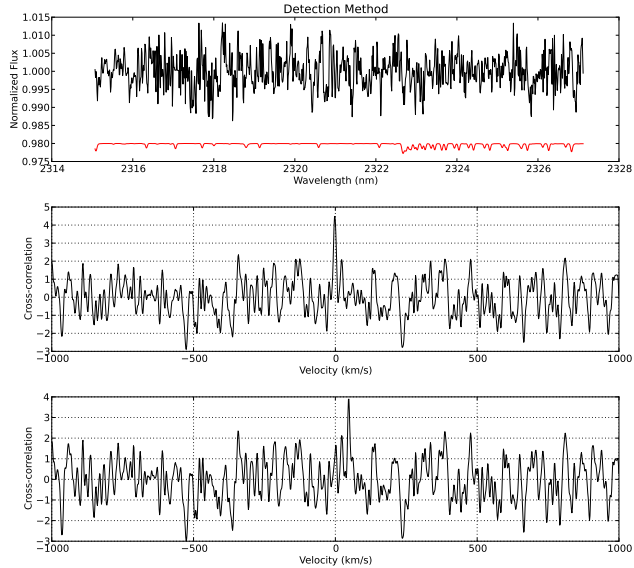}
  \caption{This figure illustrates the approximate flux ratio limit to
  the detection method outlined in section
  \ref{sec:newmethod}. \emph{Top panel}: Residuals after telluric
  correction (see section \ref{sec:reduction}) for chip 2 of HIP 80582 are in black, with an
  atmosphere model for an $0.9 M_{\odot}$ star at 50.1 Myr below it in
  red. The flux ratio at this age is $F_s/F_p = 0.0092$. \emph{Middle panel}: The scaled model
  spectrum was added to the telluric residuals, and then the sum was
  cross-correlated with the model. Despite the signal being
  significantly below the noise level, the star was detected at a high
significance. The y axis, in units of the standard deviation of the
cross-correlation function, shows that the significance of the peak is over
$4\sigma$. \emph{Bottom panel}: Same as the middle panel, but the
model spectrum was added to the residuals with a 50 km s$^{-1}$ velocity
offset.}
  \label{method}
\end{figure*}

In this paper, we introduce a technique that is sensitive to young binary
systems with secondary temperatures $4000$ K $\lesssim T_{\rm eff} \lesssim
6000 $K. For early B-type primaries with ages $\sim 15$ Myr, these
temperatures correspond to mass-ratios $q \approx 0.05-0.3$, right
where we expect to see binaries formed by disk instability (see
section \ref{sec:formation}). Rather
than attempting to detect the reflex motion of the parent
star as in exoplanet searches and SB1 binaries, we attempt to directly
detect the spectrum of the young low-mass companion using high
signal-to-noise, high-resolution data. There is a multitude of archived B-star observations in the near-infrared,
where they are used as telluric standard stars to remove the absorption spectrum of the Earth's atmosphere (telluric
lines). This method is equally sensitive to all separations within the point spread function, which is dominated by the seeing since the adaptive optics are not used in telluric standard star observations. A typical seeing of $\sim 0.8 \arcsec$ corresponds to up to $\sim 900$ AU for targets within a few kpc. In this paper, we describe a search for young F5-K9 type companions in archived VLT/CRIRES spectra of 34 early B-type stars.

In section \ref{sec:newmethod} we describe our detection method in 
more detail. Section \ref{sec:sample} describes the B-star sample we use in this work.
Section \ref{sec:reduction} contains the data reduction and telluric correction methods. 
We summarize our results in section \ref{sec:results}. We examine the
completeness of our sample in section \ref{sec:completeness} and put
limits on the multiplicity fraction as a function of mass-ratio in
section \ref{sec:multiplicity}. Finally, we present our conclusions about the prevalence of low mass-ratio companions to early B-type stars and discuss how our results constrain star formation mechanisms in Section \ref{sec:conclusions}.

\section{Direct Spectral Detection Method}
\label{sec:newmethod}

We describe here our method to detect the emission from an
approximately solar-mass star orbiting an early B-type star, which we
will hereafter call the direct spectral detection method. The basis of
this method is to cross-correlate a high signal-to-noise ratio B-star
spectrum with a synthetic F,G, or K star spectrum. If a low-mass star
with such a spectrum is orbiting the B-star, we expect to find a peak
in the cross-correlation function at the radial velocity corresponding to
the low-mass star's motion. A peak in the cross-correlation
function should appear even if the
flux from the low-mass star is comparable to or even slightly less than the noise level in the
spectrum. Figure \ref{method} illustrates the approximately limiting
case for the flux ratio. The top panel shows a fully reduced CRIRES spectrum of HIP 108975 (see section \ref{sec:reduction}) with a model spectrum for an $0.9 M_{\odot}$ star at a realistic flux ratio below it. We used
evolutionary tracks published by \cite{Landin2008} to evolve the secondary star
to 50.1 Myr, the age of the system \citep{Tetzlaff2010}, in order to determine the flux
ratio between the primary and secondary. The secondary star model was then added to
the telluric-corrected B-star spectrum at two different velocities. It is clear that the model spectrum has an amplitude much smaller
than the noise.  Nonetheless, the bottom two panels show that a cross-correlation will
have a peak with high significance at the velocity of the secondary star.

\begin{figure*}[t]
  \centering
  \includegraphics[width=\columnwidth]{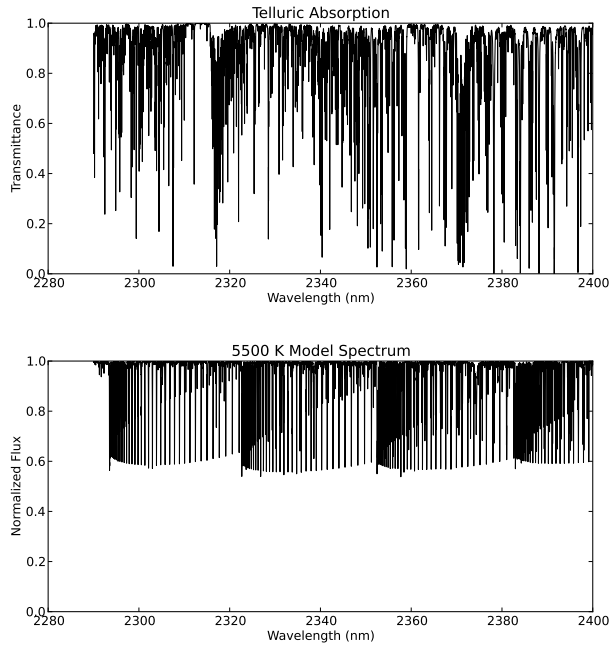}
  \caption{\emph{Top panel}: The telluric spectrum (absorption due to Earth's
    atmosphere) in the wavelength range from $2290-2400$ nm. Most of the lines are
    from CH$_4$, with a few H$_2$O lines appearing in the right
    half. \emph{Bottom panel}: The model spectrum of a 5500 K star
    with $\log (g) = 4.0$ and solar metallicity. Note that the line density of telluric
    lines is comparable to or greater than that of the star model, and many of the
    telluric lines are stronger than the stellar lines.}
  \label{telluric}
\end{figure*}

A careful choice of the wavelength region is critical for the direct spectral 
detection method. First, we want a wavelength region where the B-star spectrum
is mostly continuum (i.e. very few spectral lines). Since B-stars have few spectral lines, it is
easy to find a such a spectral region. Secondly, we want a region where the low-mass star
would have many closely spaced, strong lines. The more lines there are
in the low-mass star, the stronger the peak will be in the
cross-correlation function. Finally, we want a spectral region where the
flux ratio between the low-mass and the high-mass star is maximized.
It is not helpful to go much redder than a few microns for  
companions with $T>4000$K, because both the high-mass and
the low-mass star are firmly in the Rayleigh-Jeans limit by this
point, where the flux ratio is approximately constant. For this project,
we choose wavelengths from $2300-2400$ nm, which is the CO $\Delta\nu = 0-2$
bandhead in the low-mass star. 

There is both a lower and upper mass detection
limit. Secondary stars that are too cool will be too faint, and any
signal will be lost in the noise. Additionally, a more massive (and
hotter) primary star will decrease the flux ratio, and push the lower
mass limit up. On the other end, secondary
stars that are too hot will dissociate CO, destroying the bandhead
that we are looking for. The temperature and size of the secondary star will
depend on its age as well as its mass since it will still be
evolving towards the Main Sequence during the lifetime of the
B-star. The exact mass sensitivity will thus depend on the age,
primary star mass, and signal-to-noise ratio of the system being observed. 

The detector resolution is also important for the direct spectral detection
method. Deeper lines, providing more contrast from the
continuum, are easier to detect than broad, shallow lines. In
addition, narrow spectral lines will result in a stronger, narrower peak in the
cross-correlation function, which is most sensitive to the steep line
edges. Therefore, we want the spectral lines in the low-mass companion
to be as deep and narrow as possible. The intrinsic width of CO bandhead lines is
roughly 5-7 km s$^{-1}$. In order for the observed line width to be this
small, we need the resolution of the instrument to be $R =
\lambda/\Delta \lambda \gtrsim 50000$.

There are two main difficulties with the direct spectral detection method: 
telluric line removal and the low flux ratio between the primary and secondary
star. Figure \ref{telluric} shows the transmittance through the
Earth's atmosphere (the telluric spectrum) in the wavelength range we
are interested in. Most of the spectral lines are from methane, with a
few deep water lines towards the red end
of the range shown (See section \ref{sec:reduction} for details on the
telluric line removal). The low flux ratio makes the telluric
contamination especially troublesome, since the telluric lines are
stronger than the lines in the companion star spectrum. The flux ratio of $F_s/F_p
\sim10^{-2}$ effectively sets a lower limit on the signal-to-noise
ratio for which the direct spectral detection method is possible. 
Any flux coming from a low-mass star will be completely buried
in the Poisson noise for spectra with SNR $\ll100$. Removal of the 
telluric contamination will add more noise to the spectrum, so a spectrum
should have SNR of a few hundred \emph{before telluric line removal}
to have a good chance of detecting a companion.

\section{Star Sample}
\label{sec:sample}
B-type stars are commonly used in the near-IR as telluric
standard stars. Astronomers will observe their science targets, and
then move to a B-type star. Since B-type stars have few
spectral lines relative to cooler stars, most of the observed
spectral lines will be from the absorption of Earth's atmosphere
(telluric absorption). Therefore, these stars provide an empirical
estimate of the telluric spectrum; division of the science spectrum by
the normalized standard star spectrum will mostly remove the telluric
lines. 

Since B-type stars are commonly used as above, there are
many high signal-to-noise ratio (S/N $\gtrsim 100$)
observations of such stars in archived data. We used the VLT/CRIRES
archive in this project. CRIRES is a high
resolution ($R = \lambda / \Delta \lambda \approx 100000$) infrared
spectrograph on the VLT at Paranal Observatory. The detector consists of four
1024x512 ccd chips that are mosaiced end-to-end, and
the spectrum falls across them. There are several wavelength settings
available, which determine what parts of the spectrum fall on each 
chip. For wavelength settings in the CO
bandhead near 2300 nm, each chip will hold rougly 10 nm of spectrum
with roughly 1-2 nm gaps between the chips.

To generate the sample, we started with all single, main sequence B0-B5 stars 
with CRIRES observations from 2300-2400 nm. We then excluded any shell stars, which have circumstellar disks \citep{Porter2003} that may create
false positives. Table \ref{tab:sample} shows the complete sample used in this
project. The spectral types and ages were obtained from a catalog of nearby
young stars \citep{Tetzlaff2010}. The one exception is HIP 97611, which was
not in \cite{Tetzlaff2010}. For this star, the age was taken from
\cite{Westin1985} and the spectral type from the Simbad database\footnote{\url{http://simbad.u-strasbg.fr/simbad/}}
. The distances to all stars were
determined from parallaxes given in the Simbad database. The maximum
separation column estimates the approximate maximum separation of the binary
orbit we are sensitive to, assuming a seeing of 0.8'' which is typical of
Paranal Observatory. The median of the maximum separations to which we are sensitive is 124
AU.  The final column gives the number of distinct observations of the
star. We count all nodding positions taken on a given night with the same 
detector wavelength setting as one observation.

\begin{figure*}[ht]
  \centering
  \includegraphics[width=\columnwidth]{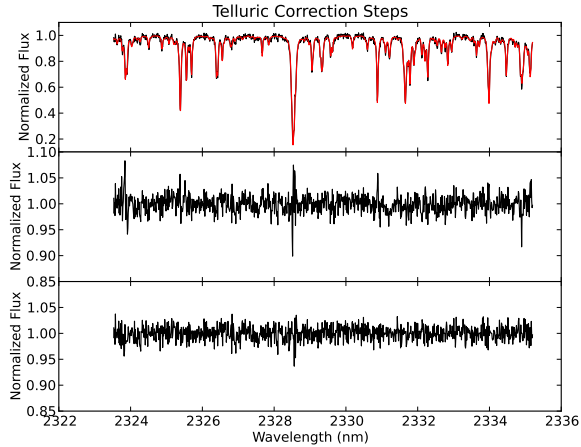}
  \caption{The telluric correction steps for chip three of CRIRES
    wavelength setting $\lambda_{ref} = 2329.3$ nm. \emph{Top panel:}
    Normalized spectrum (black), with the best-fit telluric model
    (red). \emph{Middle panel:} Residuals after dividing the observed
    spectrum by the telluric model. Note the large spikes near
    $2328.5$, $2331$, and $2335$ nm. \emph{Bottom panel:} Correction
    after fitting the large residuals to Gaussians. }
  \label{correctionsteps}
\end{figure*}

\section{Data Reduction and Telluric Correction}
\label{sec:reduction}
The data reduction was done using standard methods in
IRAF\footnote{IRAF is distributed by the National Optical Astronomy Observatories,
    which are operated by the Association of Universities for Research
    in Astronomy, Inc., under cooperative agreement with the National
    Science Foundation.}. All observations were taken in an AB or ABBA nodding
pattern. For each set of AB nods, A-B and B-A frames were made to
remove any atmospheric emission lines and dark current. The resulting
difference images were then treated to a quadratic nonlinearity correction, using
coefficients made available by the CRIRES
team%\footnote{\url{http://www.eso.org/observing/dfo/quality/CRIRES/pipeline/pipe_calib.html}}. 
The corrected frames were then divided by a normalized flat-field. Due
to the slit curvature, the spectrum can shift by up to a pixel in
the dispersion direction between the A and B nod positions. Therefore,
combining the 2D frames before extraction can reduce the spectral
resolution and affect the line shapes. For this reason, we
combined the nodding positions only after the wavelength
calibration and telluric correction. Each nod position was extracted using the 
optimal algorithm in the apall task in IRAF. The spectra were wavelength 
calibrated using a model telluric
spectrum generated with the atmospheric modeling code
LBLRTM%\footnote{\url{http://www.rtweb.aer.com/lblrtm_description.html}}
\citep{Clough2005}.

For telluric correction, we used a similar procedure to the one outlined by
\cite{Seifahrt2011}. The atmosphere modeling code
LBLRTM was used to generate a synthetic telluric
absorption spectrum. The abundances of water,
methane, and carbon monoxide were fit using a Python implementation of
a Levenberg-Marquardt nonlinear least squares fitting algorithm. The
Levenberg-Marquardt fit also refined the wavelength solution to the
telluric model, fit the continuum, and fit the resolution of the
spectrograph with a Gaussian profile. The FWHM of the profile was the
only free parameter in the resolution fit.

The LBLRTM code expects a model atmosphere, which contains the
temperature, pressure, and abundance of 30 molecules as a
function of atmospheric height. For the majority of molecular species,
we used a mid-latitude nighttime
MIPAS\footnote{\url{http://www-atm.physics.ox.ac.uk/RFM/atm/} }
profile, which provides the temperature, pressure, and abundances of
various molecules in 1 km intervals from sea level to 120 km. The
low-altitude ($z \lesssim 30$ km)
temperature, pressure \citep{Kerber2010}, and humidity
\citep{Chac2010} profiles were obtained from radiosonde data taken from Paranal Observatory.

The LBLRTM atmospheric modeling code comes with a molecular line list
based on the HITRAN 2008 database \citep{Rothman2009}, with a few
molecules individually updated. Since none of these updates were
relevant for the wavelength range from $2300 - 2400$ nm, we in
essence used the stock HITRAN 2008 database. However,
in the process of modeling, we found several water and methane lines
that were consistently under- or over-fit. For these cases, we
manually adjusted the line strengths in the database. The line
strengths were fit visually and should not be considered rigourous new line strengths. Table \ref{tab:linelist}
summarizes these changes.

In some of the 2007 data, the first chip was not well illuminated
by the flatfield lamp. This introduced an unphysical continuum shape
in the data and made the resulting model fit very poor. For these
cases, we ignored the first chip in further analysis. In addition,
the fourth chip has several bad pixels on the left edge and a streak
down the middle. None of the telluric model fits were very good on
this chip, and so we have ignored it completely in our analysis.

After the observed spectrum was fit, we found that the residuals still
contained large spikes, even on the good detector chips. These spikes
can come from a variety of sources. For the deepest lines, simple
Poisson noise can create large residuals when dividing by the telluric
model. In addition, a poorly fit continuum may cause the model to over- or under-estimate
the abundance of a given molecule. This can be especially troublesome
for water lines, for which only a few exist in the wavelength region we
are investigating. If a strong water line is near the edge of the
chip, where the continuum is usually least certain, the best-fit water
abundance may be skewed and cause none of the water lines to be well fit. We do know
that large residuals are \emph{not} coming from the spectrum of a
low-mass star, due to the expected flux ratio between the primary and
secondary, $F_s/F_p \sim10^{-2}$. Any residuals with amplitude greater than 
$1\%$ of the continuum level come from uncorrected telluric lines,
cosmic rays, or bad pixels. 

In order to minimize these spikes, we performed a second fit to any
residuals significantly above the continuum noise level. To make sure
we were not fitting away any low-mass star lines, we only corrected
spikes whose amplitude was greater than $5\%$ of the continuum level. In this
second fit, we first attempted to fit a Gaussian to each spike. If the
spike was well fit by a Gaussian, we divided the residuals by the fit.
If not, we
simply masked out the line core, so that it would not affect the
cross-correlation in later analysis (see section \ref{sec:newmethod}). Figure
\ref{correctionsteps} shows the steps involved in the telluric
correction. Notice that the secondary correction removes the
large residuals, while leaving the rest of the spectrum
unaffected.

The telluric correction described above usually reduced any telluric
lines to near the Poisson noise level in the spectrum, which is the best a
fitting routine can do. To search for any systematic errors in the
telluric correction, we added all spectra of each wavelength setting
together to make a series of master telluric residual spectra. These
master spectra had less random noise than any individual observation, and therefore we were
immediately able to see whether some telluric lines are systematically
under- or over-fit. We found that for wavelength settings with at
least ten spectra in our sample, dividing the corrected spectra by
this systematic telluric residual template increased the sensitivity
to companions. We did not make this final correction for
wavelength settings with fewer than ten individual spectra in our sample.

\section{Results}
\label{sec:results}
Each telluric residual spectrum was cross-correlated against a suite of model atmospheres
generated by the Phoenix stellar atmosphere code \citep{Hauschildt1999}. All model spectra had solar metallicity. The effective
temperatures ranged from $3000-7200$ K, in 100K intervals. We used
several surface gravities based on the stellar temperature. For the
model secondary stars with $3000 < T_{\rm eff} < 3600$, which would
have to be very young (and large) to be detectable, we used a $\log (g) = 3.5$. For the secondaries with
$3600 < T_{\rm eff} < 6500$, we used $\log (g) = 4.0$. Finally, we used
$\log (g) = 4.5$ for $T_{\rm eff} > 6500$, which can be detected
closer to the Main Sequence. We found that the surface gravity has only a very small
effect on the cross-correlation, which is more sensitive to the line
position than its precise width or depth. We
compiled a list of all cross-correlations that show a single peak with
at least $3\sigma$ significance. For a given telluric-corrected residual
spectrum, several model atmospheres may generate a significant peak at
the same velocity. This is because the model spectra of two stars
differing by only a few hundred kelvin are not very different. To keep from
counting peaks twice, we only counted the cross-correlation that
resulted in the most significant peak at a given velocity. We then attempted to
reject spurious peaks caused by the noise or incomplete telluric line
removal in a multi-stage process.

The first rejection stage was done by identifying peaks in the
cross-correlation caused by telluric residuals. To do this, we
cross-correlated a spectrum uncorrected for telluric absorption with
the same suite of model atmospheres as we used for the corrected
spectra (see above). We did these cross-correlations for one observation of each wavelength
setting. The cross-correlation of uncorrected spectra with model-atmosphere spectra generated a series of cross-correlations with peaks
arising exclusively from telluric lines. We visually compared all
of the binary candidate signals with these telluric
cross-correlations. If the dominant cross-correlation peak was at the
same velocity and had a similar width as a peak in the telluric
cross-correlation function corresponding to the same wavelength
setting and secondary model temperature, we assumed that the peak was caused by incomplete
telluric removal and rejected the candidate. There were several
cross-correlations with peaks at the same location as a telluric peak,
but with a different width. In these cases, we marked the candidate as
probably coming from incomplete telluric correction, but did not
reject the candidate.

\begin{figure*}[ht]
  \centering
  \includegraphics[width=\columnwidth]{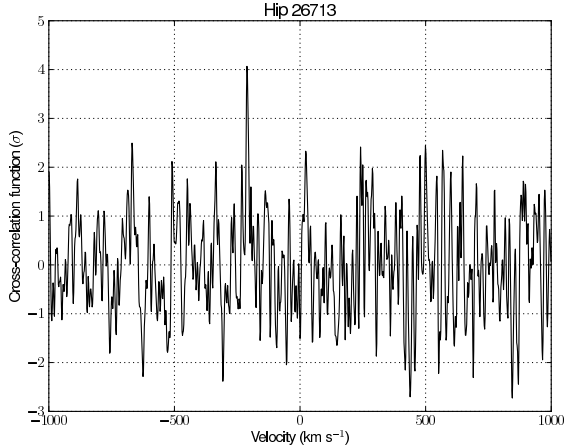}
  \caption{Cross-correlation for HIP 26713, using a 5600 K star model spectrum as template. The y-axis is in units of
    the standard deviation of the cross-correlation function. The peak is very near the maximum velocity of $|v_{\rm max}| = 256$ km s$^{-1}$, assuming a circular orbit (see equation \ref{eqn:vmax}). The likelihood of observing the system nearly edge on and at a quadrature point, so that $|v| \sim |v_{\rm max}|$, is $p \approx 0.01$. However, it is possible that the system has an eccentric orbit, effectively increasing $|v_{\rm max}|$.}
  \label{fig:hip26713}
\end{figure*}

Next, we determined whether the signal-to-noise ratio and telluric line removal in a given observation would allow us to detect the candidate companion star.
To do this, we added a model atmosphere with the
same temperature as the candidate to the telluric-corrected spectrum at 17 different radial
velocities ranging from -400 to 400 km s$^{-1}$. We do not expect to see any
peaks from real companions with $|v| > 400$ km s$^{-1}$, the approximate
radial velocity of a $1 M_{\odot}$ star orbiting a $10 M_{\odot}$ star
such that the stellar surfaces are in contact. The flux ratio of the model atmosphere to the primary was obtained by
interpolating pre-main-sequence evolutionary tracks from
\cite{Landin2008} at the age of the system, as well at age$\pm
\sigma_{age}$. The primary star ages for our sample are given in Table
\ref{tab:sample}. We cross-correlated each of these semi-synthetic
spectra against the model spectrum; if the largest
peak was at the correct velocity, we counted the star as
detected. If the star was not detected in the sensitivity
analysis at least $50\%$ of the time, we rejected the candidate.
The significance of the correct peak in the cross-correlation 
function can vary greatly, depending on where the stellar spectrum falls in 
relation to the telluric line residuals. Therefore, we cannot usually reject a peak
based solely on its significance.

We then visually inspected the remaining candidate cross-correlations,
picking out those with a single dominant peak with $|v_r| < v_{\rm max}$
where $v_{\rm max}$ is the maximum possible radial velocity for a star of
temperature $T_{\rm sec}$ to be orbiting a hotter star of temperature
$T_{\rm prim}$ with a semimajor axis $a$. Assuming a circular orbit,
$v_{\rm max}$ is given by

\begin{equation}
v_{\rm max} = \sqrt{\frac{2G(M_{\rm prim} + M_{\rm sec})}{R_{\rm prim}}} \cdot \frac{T_{\rm sec}}{T_{\rm prim}}
\label{eqn:vmax}
\end{equation}

where $M_{\rm prim}$ and $M_{\rm sec}$ are the masses of the primary and
secondary stars, respectively, and $R_{\rm prim}$ is the radius of the
primary star. The primary masses are given in \cite{Tetzlaff2010}, while
the radii and primary star temperatures were estimated from spectral type relations given in
\cite{CarrollOstlie}. An eccentric orbit could have a larger maximum velocity than
that estimated by equation \ref{eqn:vmax}, if the orbit was oriented
such that the the secondary star was moving directly towards or away from earth at or near
periastron. A star in an eccentric orbit cannot get so close to the
primary that they touch, and its \emph{average} distance must still be
far enough to allow for the observed secondary star temperature. These
conditions lead to a maximum eccentricity, given by

\begin{equation}
e_{\rm max} = 1 - \frac{v_{\rm max}^2 (R_{\rm prim} + R_{\rm sec})}{G(M_{\rm prim} + M_{\rm sec})}
\label{eqn:emax}
\end{equation}

where $R_{\rm prim} $ and $R_{\rm sec}$ are the radii of the primary
and secondary stars, respectively, and $v_{\rm max}$ is the maximum
circular velocity given by equation \ref{eqn:vmax}. Typical values
give $e_{\rm max} \approx 0.6$. Eccentricities near this value could
allow for velocities significantly greater than $v_{\rm max}$ given by
equation \ref{eqn:vmax}.

The above analysis is summarized in Table \ref{tab:rejection}. There
are two binary candidate systems that we have not been able to reject.
 For both of these, we checked what other observations the candidate
star had within the CO bandhead spectral region ($2300-2400$ nm). The analysis of each
of these stars is done separately below.

%\newpage

\subsection{HIP 26713}
The cross-correlation for this candidate is shown in figure
\ref{fig:hip26713}. The strong peak at -220 km s$^{-1}$ has a
significance just over $4\sigma$, and corresponds to a 5600 K star
model. A sensitivity analysis (Table \ref{tab:rejection}, step 2) gives a median peak significance of $\sim 6\sigma$
with a large ($\sim 2\sigma$) spread. Due to the large spread, we cannot reject
the peak based on the observed significance.

The candidate radial velocity amplitude of 220 km s$^{-1}$ is very near the upper limit 
given by equation \ref{eqn:vmax} of 256 km s$^{-1}$. If this candidate is a real binary companion, it
must have been observed very near its radial velocity maximum and the orbital inclination must be very
near edge-on. Assuming a circular orbit and equal probabilities
of observing any given phase or inclination, the probability of this
occuring for this system is $\sim 0.01$. However, the probability for one system in our entire sample to be caught with this chance alignment rises to $0.14$.  In addition, we cannot discount the
possibility of an eccentric orbit leading to a higher value of $v_{\rm
  max}$ than estimated by equation \ref{eqn:vmax}. Without more data, we can neither confirm nor disprove the existence of a companion star orbiting HIP 26713.
  
If HIP 26713 is a true binary system, evolutionary tracks by \cite
{Landin2008} give a secondary star mass of $1.6 \pm 0.2 M_{\odot}$. 
The mass of the primary star is $9.4 \pm 0.2 M_{\odot}$ \citep
{Tetzlaff2010}, giving a mass-ratio of $q = 0.17 \pm 0.02$. For a circular orbit, the corresponds to an orbital period of 10.0 days. If the companion star is on an eccentric orbit, its true period would be longer than this.

\begin{figure*}[ht]
  \centering
  \includegraphics[width=5.7in]{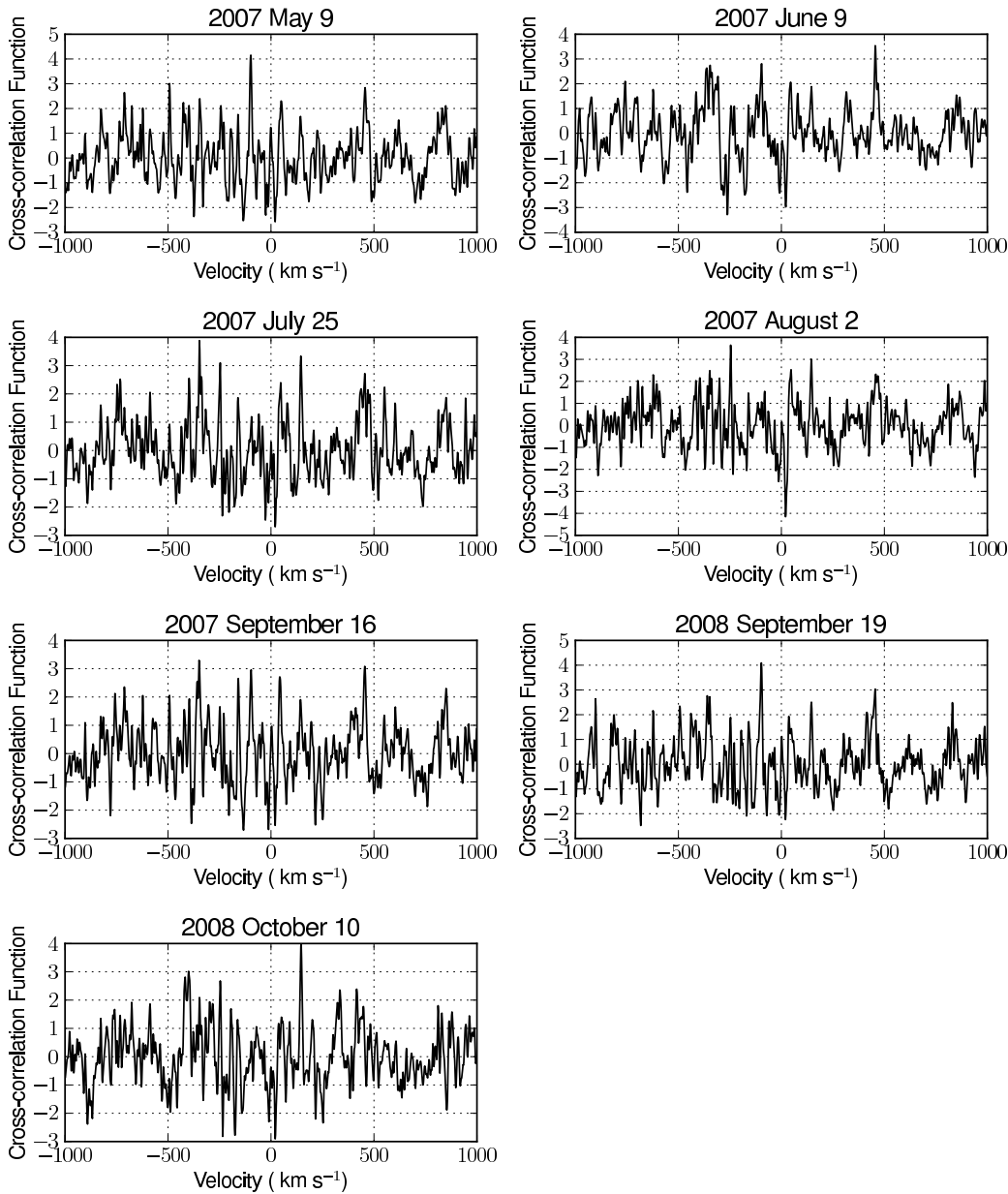}
  \caption{Cross-correlations for HIP 92855, for all dates observed. A 6100 K star model spectrum is used as the template for each cross-correlation. The y-axis is in units of
    the standard deviation of the cross-correlation function. A single strong peak is seen in the
    cross-correlations from 2007 May 9, 2007 August 2, 2008 September
    19, and 2008 October 10. The reasonably strong peak on 2007 June 9 is identified
  as probably arising from imperfect telluric line removal or random noise, since it
  has $v>v_{\rm max}$ given by equation \ref{eqn:vmax}.}
  \label{fig:hip92855}
\end{figure*}

\subsection{HIP 92855}
\label{sec:hip92855}
There were seven observations of HIP 92855 on different dates, all with the 2336 nm
wavelength setting. The cross-correlations for four of the observation dates show
a single strong peak when using a 6100K model star as template. Figure
\ref{fig:hip92855} shows the cross-correlations of the
telluric-corrected spectra with a 6100K model for all of the
observations. Table \ref{tab:hip92855} summarizes the sensitivity and
cross-correlation significance. The detection rate is the fraction of the 17
radial velocities between -400 and 400 km s$^{-1}$ that were correctly detected in the sensitivity
analysis (Table \ref{tab:rejection}, step 2). The variation in detection rate is due to the different
signal-to-noise levels and telluric line corrections in the observations
at different dates. The expected significance is the median significance of
the radial velocities which were detected in the sensitivity analysis, in units of
the standard deviation of the cross-correlation function. The observed
significance and velocity are for the observed peaks. The velocities in Table \ref{tab:hip92855} are corrected for
the barycentric motion and the known systematic radial velocity of HIP 92855,
while those in Figure \ref{fig:hip92855} are not.

As Table \ref{tab:hip92855} shows, a 6100K star orbiting HIP 92855 is at
the limit of detectability with the direct spectral detection
method. With the exception of the observation on 2007 September 16, the
cross-correlations for the dates with the highest detection rates have
a single large peak. We consider this an excellent candidate for
follow-up observations.

\begin{figure*}[ht]
  \centering
  \includegraphics[width=5.5in]{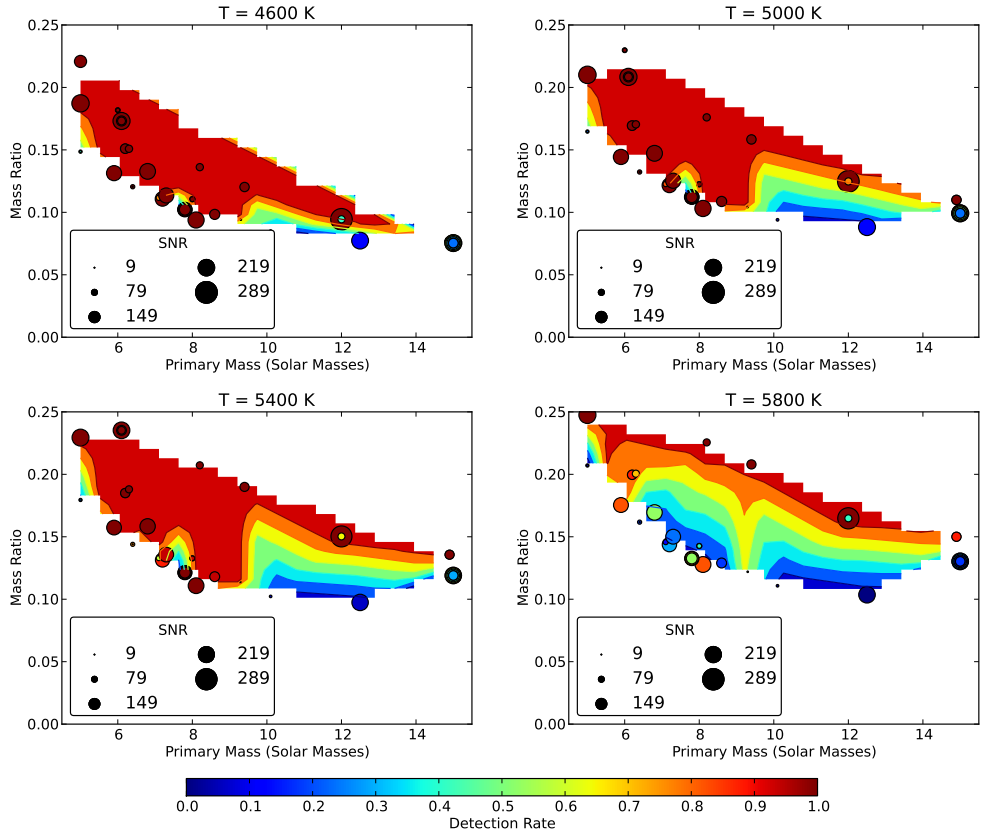}
  \caption{Completeness diagram for the full sample of main-sequence B
  stars, split up by the effective temperature of the secondary
  star. The points correspond to the individual stars in the sample,
  and their sizes reflect the signal-to-noise ratio in the
  spectrum. Note that the signal-to-noise is calculated after the
  telluric line removal, and counts any telluric residuals as
  noise. The figures are also color-coded by the fraction of trials
  that detected the companion (see section
  \ref{sec:completeness}). Contours are drawn to guide the eye. The red 
  areas in each plot indicate the regions for which our sample is complete.}
  \label{fig:completeness}
\end{figure*}

If HIP 92855 is a true binary system, evolutionary tracks by
\cite{Landin2008} give a secondary star mass of $1.2 \pm 0.2 M_{\odot}$. The mass of the
primary star is $7.8 \pm 0.2 M_{\odot}$ \citep{Tetzlaff2010}, giving a mass
ratio of $q = 0.15 \pm 0.04$. The maximum radial velocity, observed on 2007 August 2, was 234 km s$^{-1}$. If this is the radial velocity semi-amplitude and if the companion is on a circular orbit, the binary orbit would have a period of 6.8 days. If the true velocity semi-amplitude is larger either because no observation was taken when the companion star was at quadrature or because the orbit is inclined, the period would be shorter than this. If the companion star is on an eccentric orbit, the period could be longer than 6.8 days if the large velocity observed was near periastron.

\section{Completeness}
\label{sec:completeness}
We now estimate the completeness of the direct spectral detection
method applied to this data set. For each telluric-corrected observation, we
created a series of synthetic binary-star spectra by adding stellar models to
the data at various flux ratios, temperatures, and radial
velocities. We used evolutionary tracks from \cite{Landin2008} to
find the luminosity of model companion stars with temperatures ranging from 3000K to
7000K in steps of 500 K. To find the model flux ratio $F_s/F_p$, we
used the best-fit age for each star quoted in
Table \ref{tab:sample}, as well as the best-fit age$\pm
\sigma_{age}$. Model secondary spectra generated with the Phoenix code
\citep{Hauschildt1999} were added to the telluric-corrected
observations at 17 different radial velocities ranging
from -400 to +400 km s$^{-1}$. Changing the radial velocity of the model
spectrum changes where the companion spectral lines fall with respect
to the telluric lines. Finally, we cross-correlated each synthetic
spectrum with its corresponding model secondary star spectrum, and examined the cross
correlation function. 

If the highest peak in the cross-correlation function was at the
correct velocity, the companion was considered detected. We then
tabulated how many times the companion star was detected in the 17
radial velocity trials. Figure \ref{fig:completeness} shows the
fraction of trials that detected the companion for all of the model
radial velocities, as a function of primary (B-star) mass and the binary mass-ratio. The points
correspond to individual spectra, with their sizes indicating the
signal-to-noise ratio in the spectrum, and the contours are drawn by
interpolating between the points.  The four panels are for different companion star temperatures. Companion stars with effective
temperatures from $4600-5400$ K have large regions with a very high
detection rate. Stars cooler than about 4600 K are too dim to detect
without much higher signal-to-noise ratios than present in our dataset, and stars hotter than about
5400 K do not have a strong CO bandhead and so the cross-correlation
function is not as sensitive. 

Figure \ref{fig:completeness} shows that the direct spectral detection
method is able to find companion stars with a mass-ratio of $q\approx
0.1-0.2$, for a range of effective temperatures. The regions with a 
detection rate near 1 are completely sampled, and a companion star in 
that region would be detected. For primary stars with $M < 10 M_{\odot}$, we are sensitive to almost all companions with $4600 < T_{\rm eff} < 5400$ K.

\begin{figure*}[ht]
  \centering
  \includegraphics[width=\columnwidth]{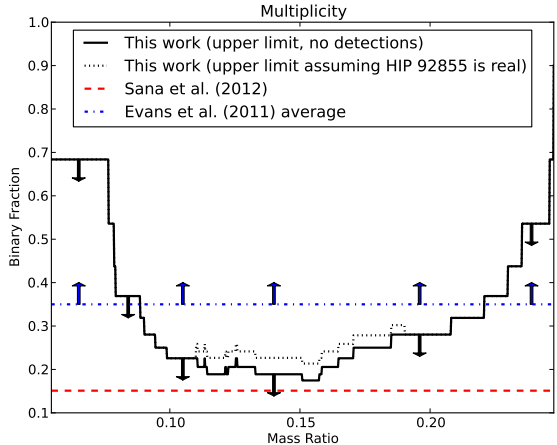}
  \caption{Estimates of the binary fraction of B0-B5 stars,
    as a function of binary mass-ratio. This work found no unambiguous companions,
    and so we give $90\%$ upper limits (solid black line). $90\%$
    upper limits are also given assuming that HIP 92855 is a real
    binary system (dotted black line). The upper limits are only different within the
    $1\sigma$ error bars on the mass-ratio for HIP 92855. The nearly flat 
    distribution found by \cite{Sana2012} is shown as the dashed red line. The average binary 
    fraction found by \cite{Evans2011} is also shown (dash-dot blue line). The
    \cite{Evans2011} value is a \emph{lower} limit and an average over all mass-ratios from $0.1 < q < 0.3$, but they estimate
    that their sample is very complete, and so the true multiplicity fraction is quite close to their value.}
  \label{fig:limits}
\end{figure*}

\section{Multiplicity Fraction}
\label{sec:multiplicity}

We have not found any unambiguous low mass-ratio companions in our
sample, though we do have two candidates that require follow-up
observations (HIP 92855 amd HIP 26713). From the work described in section
\ref{sec:completeness}, we define a range
of mass-ratios for which the direct spectral detection method is
sensitive for each primary (B-) star. We can then rule out any
companions with mass-ratios in that range, for that primary star.

In order to convert these star-by-star limits on the presence of a
companion into upper limits on the multiplicity
fraction of the parent population, we first count the number of stars
that rule out companions in a particular range of mass-ratios. We then
apply binomial statistics, where the probability P of finding k
companions from n samples of a parent population with a true binary
fraction p is given by

\begin{equation}
P(k|p,n) = \frac{n!}{k!(n-k)!}p^k(1-p)^{n-k}
\label{eqn:binomial}
\end{equation}

For no detected binary companions (k=0), the corresponding likelihood
function for the binary fraction is

\begin{equation}
P(p|k=0, n) = (n+1)\cdot  (1-p)^n
\label{eqn:likelihood}
\end{equation}

A $90\%$ upper limit is given by

\begin{equation}
0.9 = \int_0^p (n+1)\cdot  (1-p')^n dp'
\label{eqn:limitdef}
\end{equation}

with solution

\begin{equation}
p_{90} = 1 - 0.1^{\frac{1}{n+1}}
\label{eqn:limit}
\end{equation}

A similar derivation gives $90\%$ upper limits for one detection (k=1).
Figure \ref{fig:limits} shows the $90\%$ upper limits to the binary
fraction as a function of mass-ratio. To find n in equation
\ref{eqn:limit}, we counted the number of stars that ruled out a
companion at the given mass-ratio. We only counted companions that
were found in all 17 radial velocity trials (see section
\ref{sec:completeness}), and were always found with at least $4\sigma$
significance. We also included upper limits assuming that HIP 92855, the more likely of our two candidates, is
a true binary system. Figure \ref{fig:limits} also shows the lower
multiplicity limit set by \cite{Evans2011} (blue dotted line) and the
\emph{intrinsic} O star mass-ratio distribution derived by
\cite{Sana2012}. \cite{Evans2011} do not split their multiplicity by mass-ratio,
and so the line shown in figure \ref{fig:limits} is an average value. While
we include them for comparison with our results, neither of these studies are
directly comparable to our sample. \cite{Sana2012} have only O stars in
their sample and are only able to measure mass-ratios $q\approx
0.2-1$, although they consider mass-ratios down to $q=0.1$
when deriving the intrinsic distribution. \cite{Evans2011} are sensitive to similar mass-ratios as our
sample, but they sample late B-stars (B4-B9) while we sample early B
stars (B0-B5). Our results are almost perfectly complementary to those of \cite{Evans2011}. It
is encouraging that our upper limits for $0.1<q<0.2$, where our sample
is most complete, lie in between the results of
\cite{Sana2012} who measure the binary fraction of more massive
primaries, and \cite{Evans2011} who measure the binary fraction of
less massive primaries than we do.

The above analysis assumes that the sample of B-stars given in Table
\ref{tab:sample} is representative of the B-star population as a
whole. Most of the sample stars are field
B-stars, which have a lower overall multiplicity than cluster or
association B-stars \citep{Mason2009}. However, the close binaries this
method is sensitive to would be difficult to disrupt with dynamical
interactions in a cluster environment, and we therefore expect this
sample to be representative of both populations. A potential complication is that telluric standard stars are often chosen specifically because they are \emph{not} known binary systems. While this may introduce a bias for large mass-ratios, the direct spectral detection method is sensitive to low mass-ratio companions that may not have been found with other methods such as classical spectroscopy or imaging and so companions with $q \lesssim 0.2$ may be less effected. The amount of bias introduced into the above measurement depends on how carefully the individual observers chose the telluric standard stars, and so is very difficult to assess.

\section{Conclusion}
\label{sec:conclusions}
We have described a new technique for finding binary systems with a
flux ratio of $F_p/F_s \approx 100$, where $F_p$ and $F_s$ are
the fluxes from the primary and secondary star, respectively. In this
technique, which we call the direct spectral detection technique, we
use high signal-to-noise, high resolution spectra of a binary candidate.
We remove the contamination from the Earth's atmosphere with the telluric 
modeling code LBLRTM, and cross-correlate the residuals with a library of stellar 
models for late type stars (F2-M5). A binary detection would appear as a strong 
peak in the cross-correlation function.

We prove the feasibility of the direct spectral detection method by
adding a synthetic signal to real data, and successfully recovering
it. We further investigate the completeness of the method in section \ref{sec:completeness}. This method is sensitive to detecting a range of companion stars, set by the spectral type and age of the primary and the signal-to-noise ratio of the observation. Our sample is sensitive to almost all companion stars with $4600 < T < 5400$, corresponding to binary mass-ratios of $0.1 \lesssim q \lesssim 0.2$.

 We have applied this technique to a sample of 34 archived main sequence 
early B-stars (B0-B5) with spectra taken with the CRIRES near-infrared
spectrograph. We found no unambiguous companions in our sample, but identify two targets as candidate binary systems: HIP 92855 and HIP 26713. HIP 92855 is B2.5V type star, with a candidate companion star with effective temperature $T = 6100$ K and mass $1.2 \pm 0.2 M_{\odot}$. Such a companion star is very near the detection limit of the direct spectral detection technique and deserves further follow-up observations. HIP 26713 is a B1.5V type star, with a candidate companion star with $T = 5600$ K and $M = 1.6 \pm 0.2 M_{\odot}$. This star was only observed once in our sample, and so may be a series of incompletely removed telluric absorption lines masquerading as a companion star.

We set upper limits on the binary fraction of early B stars as a function of binary mass-ratio (see Figure \ref{fig:limits}). As well as showing the upper limit for no detections in our sample, we also show the binary fraction upper limit assuming the HIP 92855 is a true binary system.  Our upper limits are strongest for mass-ratios $q \approx 0.1 - 0.15$, and are
about $20\%$. We compare our limits to the intrinsic binary mass-ratio distribution for O-type primaries derived by \cite{Sana2012}, as well as the lower limit \emph{average} binary fraction seen by \cite{Evans2011} for late B-stars (B4-B9). Our strongest upper limits ($0.1 \leq q \leq 0.15$) fall in between these two previous studies.

Companion stars formed by circumstellar disk instability are expected to have
typical mass-ratios near $q = 0.1$ \citep{Kratter2006,
  Stamatellos2011}, near where our upper limits are strongest. If there was
a large population of low mass-ratio companions formed by disk
fragmentation, we would expect to see a peak in the mass-ratio
distribution near $q \approx 0.1$. Since our results agree well with
the nearly flat mass-ratio distribution derived by \cite{Sana2012}, it
is unlikely that such a peak exists. There are several possible
interpretations of this result, three of which we give below.

\begin{enumerate}
\item Stellar companions formed by disk instability have a much lower
  characteristic mass-ratio than $q=0.1$ which remain invisible to
  observations.
\item The mass-ratio distribution of companions formed by disk
  instability is very broad, and so we would not expect a strong peak
  in the observed mass-ratio distribution. This interpretation may be
  supported by the flatness of the mass-ratio distribution as well as
  disk instability simulations that end with massive companions
  \citep[e.g.][]{Krumholz2009, Clarke2009}.
\item Disk instability is not a dominant formation mechanism for low
  mass-ratio binary systems, and molecular core fragmentation alone
  can generate the nearly flat mass-ratio distribution down to low mass-ratios.
\end{enumerate}

It is difficult to distinguish between these interpretations at this
time. More observational work, as well as computational work, is
required to explain the binary properties of high-mass stars. 

We would like to acknowledge Andreas Seifahrt, Rob Robinson, and Daniel Jaffe for their generous help with the telluric modeling 
procedure and some of the statistical aspects of this work. We would
also like to thank the referee for a quick review and many helpful comments. This 
research has made use of the 
following online resources: the SIMBAD database and the VizieR
catalogue access tool at CDS, Strasbourg, France, and the ESO Science 
Archive Facility. Funding for this work was provided by a National Science Foundation CAREER award to Sarah Dodson-Robinson (AST-1055910) and by start-up funding from the University of Texas College of Natural Sciences.

%%%%%%%%%%%%%%%%%%%%%%%%%%%%%%%%%%%%%%%%%%%%%%%%%%%%%%%%%
%%                Tables                           %%%%%%
%%%%%%%%%%%%%%%%%%%%%%%%%%%%%%%%%%%%%%%%%%%%%%%%%%%%%%%%%

%\begin{center}
\begin{small}
\begin{longtable*}{|cccccc|}
    %\tablecaption{Full star sample}
    %\tablefirsthead{\hline Star & Spectral Type & Age (Myr) & Distance
      %(pc) & Maximum Separation (AU) \\ \hline} 

    %\tablehead{\multicolumn{5}{c}{{\tablename} \thetable{} -- Continued} \\ \hline Star &
     % Spectral Type & Age (Myr) & Distance (pc) & Maximum Separation (AU) \\ \hline} 

    %\tabletail{\hline}

   % \tablelasttail{\hline}

    \caption{Full star sample} \\
        \hline
       & & & & Maximum & Number of \\ Star & Spectral Type & Age (Myr) & Distance
       (pc) & Separation (AU) & Observations \\ \hline
        \endfirsthead

        \multicolumn{6}{c}{{\tablename} \thetable{} -- Continued} \\
        \hline
         & & & & Maximum & Number of \\ Star & Spectral Type & Age (Myr) & Distance
       (pc) & Separation (AU) & Observations \\ \hline
        \endhead

        \hline
        \endfoot

        \hline
        \endlastfoot

%    \begin{supertabular}{| c c c c c |}
%        HIP 108975 & B3V (+B) 	&$ 50.1 \pm 10.9 $& 19.96 & 15.97 & 2 \\ 
        HIP 23364 & B3V &$ 31.6 \pm 0.6 $& 31.65 & 25.32 & 1 \\ 
        HIP 26713 & B1.5V &$ 7.2 \pm 2.5 $& 138.89 & 111.11 & 1 \\ 
        HIP 27204 & B1IV/V & $12.6 \pm 4.6$ & 408.2 & 326.5 & 1 \\
	HIP 30122 & B2.5V & $32 \pm 0.4$ & 111.1 & 88.9 & 4 \\
	HIP 32292 & B2V & $8.2 \pm 0.1$ & 1111.1 & 888.9 & 1 \\
	HIP 39866 & B3V & $25.1 \pm 2.6$ & 840.3 & 672.3 & 1 \\
	HIP 48782 & B3V & $32.3 \pm 0.6$ & 370.4 & 296.3 & 1 \\
        HIP 52370 & B3V &$ 17.2 \pm 1.3 $& 58.14 & 46.51 & 2 \\ 
        HIP 52419 & B0Vp &$ 4 \pm 0.7 $& 250 & 200 & 2 \\ 
	HIP 54327 & B2V & $11.7 \pm 6.2$ & 252.5 & 202.0 & 3 \\
	HIP 55667 & B2IV-V & $22.5 \pm 2.6$ & 847.5 & 678.0 & 1 \\
        HIP 60823 & B3V	&$ 25.3 \pm 6.3 $& 39.53 & 31.62 & 5 \\ 
        HIP 62327 & B3V &$ 8.2 \pm 1.8 $& 121.95 & 97.56 & 4 \\
        HIP 63945 & B5V &$ 27.3 \pm 11.4 $& 36.63 & 29.3 & 1 \\  
   	HIP 61585 & B2IV-V & $18.3 \pm 3.2$ & 96.7 & 77.4 & 8 \\
   	HIP 62327 & B3V & $8.2 \pm 1.8$ & 117.9 & 94.3 & 4 \\
	HIP 63007 & B4Vne & $53.3 \pm 8.1$ & 117.6 & 94.1 & 2 \\
	HIP 63945 & B5V & $27.3 \pm 11.4$ & 119.6 & 95.7 & 1 \\
	HIP 67796 & B2V	& $15.4 \pm 0.4$ & 970.9 & 776.7 & 1 \\
        HIP 68282 & B2IV-V &$ 13 \pm 2 $& 76.92 & 61.54 & 2\\ 
        HIP 68862 & B2V &$ 9.1 \pm 3.8 $& 109.89 & 87.91 & 1 \\ 
        HIP 71352 & B1Vn + A &$ 5.6 \pm 1 $& 178.57 & 142.86 & 2 \\ 
        HIP 73129 & B4Vnpe &$ 27.1 \pm 6.1 $& 36.9 & 29.52 & 1 \\ 
        HIP 74110 & B3V &$ 33.2 \pm 7.3 $& 30.12 & 24.1 & 1 \\ 
        HIP 76126 & B3V &$ 15.9 \pm 1.3 $& 62.89 & 50.31 & 1 \\ 
        HIP 78820 & B0.5V & $13.8 \pm 0.4$ & 123.9 & 99.1 & 1 \\
        HIP 80582 & B4V &$ 50.1 \pm 14 $& 19.96 & 15.97 & 2 \\ 
        HIP 80815 & B3V &$ 10.5 \pm 2.1 $& 95.24 & 76.19 & 4 \\ 
        HIP 81266 & B0.2V &$ 5.7 \pm 1 $& 175.44 & 140.35 & 12 \\ 
        HIP 82514 & B1.5Vp+ &$ 20 \pm 2 $& 50 & 40 & 1 \\ 
        HIP 87314 & B2/B3Vnn &$ 23.2 \pm 2.9 $& 43.1 & 34.48 & 7 \\ 
        HIP 92855 & B2.5V &$ 31.4 \pm 0.4 $& 31.85 & 25.48 & 7 \\ 
        HIP 92989 & B3V	&$ 7.9 \pm 2.1 $& 126.58 & 101.27 & 1 \\ 
%        HIP 94385 & B3V &$ 27.9 \pm 4.1 $& 35.84 & 28.67 & 2 \\
        HIP 97611 & B5V &$ 45 \pm 10 $& 66.67 & 53.33 & 1        
 %       \tablecomments{The maximum
%separation column estimates the approximate separation of the binary
%orbit we are sensitive to, assuming a seeing of 0.8''}
%      \end{supertabular}

        \label{tab:sample}
\end{longtable*}
\end{small}
%\end{center}

\newpage
%\begin{center}
   \begin{longtable*}{|cccc|}
      \caption{Summary of adjusted line strengths. The units of line
        strength are cm$^{-1}$/(molecule $\times$ cm$^{-2}$)} \\
        \hline
        Wavelength (nm) & Molecule & Old Strength & New Strength \\ \hline \hline
        \endfirsthead

        \multicolumn{4}{c}{{\tablename} \thetable{} -- Continued} \\
        \hline
        Wavelength & Molecule & Old Strength & New Strength \\ \hline \hline
        \endhead

        \hline
        \endfoot

        \hline
        \endlastfoot

        2317.12   &   CH$_4$   &   5.445\e{-21}   &   5.034\e{-21} \\
        2318.24   &   H$_2$O   &   1.400\e{-24}   &   2.256\e{-24} \\
        2328.51   &   CH$_4$   &   2.521\e{-21}   &   2.371\e{-21} \\
        2328.56   &   CH$_4$   &   1.270\e{-21}   &   1.358\e{-21} \\
        2340.12   &   CH$_4$   &   3.085\e{-21}   &   2.963\e{-21} \\
        2340.36   &   CH$_4$   &   3.343\e{-21}   &   3.211\e{-21} \\
        2351.64   &   H$_2$O   &   1.670\e{-23}   &   1.393\e{-23} \\
        2351.69   &   H$_2$O   &   1.085\e{-23}   &   7.985\e{-24} \\
        2352.43   &   CH$_4$   &   3.144\e{-23}   &   4.031\e{-24} \\
        2352.45   &   H$_2$O   &   4.639\e{-23}   &   4.939\e{-23} \\
        2353.62   &   CH$_4$   &   2.708\e{-21}   &   2.654\e{-21} \\
        2355.82   &   CH$_4$   &   5.101\e{-21}   &   4.949\e{-21} \\
        2358.9     &   CH$_4$   &   5.160\e{-21}   &   4.710\e{-21} \\
        2364.03   &   H$_2$O   &   1.408\e{-23}   &   1.217\e{-23} \\
        2367.23   &   H$_2$O   &   2.078\e{-23}   &   2.182\e{-23} \\
        2370.35   &   CH$_4$   &   4.028\e{-21}   &   3.625\e{-21} \\
        2370.41   &   CH$_4$   &   2.437\e{-21}   &   2.021\e{-21} \\
        2370.75   &   CH$_4$   &   1.466\e{-21}   &   9.138\e{-22} \\
        2371.39   &   H$_2$O   &   3.905\e{-23}   &   3.171\e{-23} \\
        236.62     &   H$_2$O   &   1.146\e{-23}   &   9.186\e{-24} \\
        2376.63   &   H$_2$O   &   3.824\e{-24}   &   3.820\e{-24} \\
        2378.2     &   H$_2$O   &   1.134\e{-22}   &   1.021\e{-22} \\
        2379.67   &   H$_2$O   &   6.334\e{-24}   &   8.408\e{-24} \\
        2385.98   &   H$_2$O   &   6.051\e{-23}   &   5.407\e{-23}

    \label{tab:linelist}
  \end{longtable*}
%\end{center}

%\begin{center}
 \begin{small}
   \begin{longtable*}{|ccp{6cm}|}
      \caption{Summary of Cross-Correlation Function Peak Rejection Steps. See Section \ref{sec:results} for more information} \\
        \hline
        Step  & Description & Method \\ \hline
        \endfirsthead

        \multicolumn{3}{c}{{\tablename} \thetable{} -- Continued} \\
        \hline
        Step  & Description & Method \\ \hline
        \endhead

        \hline
        \endfoot

        \hline
        \endlastfoot

        1 & Telluric Residual Peak Identification & Compare cross-correlation function of telluric-corrected spectrum with that of an uncorrected spectrum. \\

        \hline
        2 & Sensitivity Analysis & Check that signal-to-noise ratio
        is high enough to detect secondary candidate at a range of velocities \\

        \hline
        3 & Velocity Analysis & Check that a blackbody with the
        candidate temperature can exist as close to the primary B-star as the velocity indicates (assumes circular orbit)

    \label{tab:rejection}
  \end{longtable*}
 \end{small}
%\end{center}

\scriptsize 
\begin{samepage}
%\begin{center}
\begin{longtable*}[t]{|lcccc|}

\caption{Summary of HIP 92855 observations. Significance is in units of
  the standard deviation of the cross-correlation function. The detection rate and expected significances are from the sensitivity analysis (Table \ref{tab:rejection}, step 2).} \\
\hline
Date & Detection Rate & Expected Significance & Observed Significance
& Velocity (km s$^{-1}$) \\ \hline
\endfirsthead

\multicolumn{4}{c}{{\tablename} \thetable{} -- Continued} \\
\hline
Date & Detection Rate & Expected Significance & Observed Significance
& Velocity (km s$^{-1}$) \\ \hline
\endhead

\hline
\endfoot

\hline
\endlastfoot

2007 May 9 & 0.35 & 3.3 $\sigma$ & 4.1 $\sigma$ & -74 \\
2007 June 9 & 0.18 & 3.6 $\sigma$ & N/A & N/A \\
2007 July 25 & 0.35 & 3.7 $\sigma$ & N/A & N/A \\
2007 August 2 & 0.47 & 3.6 $\sigma$ & 3.5 $\sigma$ & -234 \\
2007 September 16 & 0.82 & 4.2 $\sigma$ & N/A & N/A \\
2008 September 19 & 0.71 & 3.4 $\sigma$ & 4.1 $\sigma$ & -126 \\
2008 September 10 & 0.59 & 3.3 $\sigma$ & 4.0 $\sigma$ & 116 
\label{tab:hip92855}
\end{longtable*}
%\end{center}  
\end{samepage}
%Return to normal font size
\normalsize

\newpage 

%\bibliography{references}

\end{document}